# Low voltage graphene interface engineered organic ferroelectric tunnel junction devices

S. Natani[1], P. Khajanji[2], L. Cheng[1], K. Eshraghi[2], Z. Zhang[3], W. Shipley[2], A. Tao[2] and P.R. Bandaru[1,2]

[1]Department of Mechanical Engineering, University of California, San Diego, La Jolla, CA

[2]Program in Materials Science, University of California, San Diego, La Jolla, CA

[2]Applied Materials, Santa Clara, CA

**Abstract**: It has been indicated [1] that the path forward for the widespread usage of ferroelectric (*FE*) materials may be considerably facilitated through the reduction of programming voltages to on-chip logic compatible values of < 1 V. Obstacles involve issues related to the scaling of the *FE*s to lower thickness as well as the presence of an interfacial layer (IL) between the high permittivity *FE* [2] and the substrate- resulting in *wasted* voltage across the IL. Here, we show how lower operating voltages along with a higher tunneling electroresistance (TER) could be achieved through IL engineering. We use piezoresponse force microscopy and fabricated ferroelectric tunnel junctions (FTJs)[3] to show that ultra-thin FE films deposited on single layer graphene can exhibit polarization switching at ~ 0.8 V with significant TER.

**Introduction.** FTJs have been mostly constructed using *distinct* metal (*M*) electrodes on either side of the *FE*, *i.e.*, in a *M1/FE/M2* arrangement: **Fig. 1(a)** – so as to harness the difference in the tunneling currents through the *FE* with respect to polarization direction[5]. The bound charges corresponding to a given *FE* polarization ($P_{FE}$), are compensated over a screening length ($l_{scr}$) in the electrodes. The distinction in the respective $l_{scr}$ between the two metals implies directionality dependent barriers to current transmission yielding low(/high) resistance state current/s, termed as $I_{LRS}$ (/ $I_{HRS}$). The fidelity between the two states is manifested through the magnitude of the tunneling electroresistance: $TER\ (\%) = \frac{I_{LRS} - I_{HRS}}{I_{HRS}} \times 100$. The *TER* is then typically[6] proportional to the relative barrier height differences, and would be further enhanced through using a semiconductor (*S*) as one of the electrodes, *i.e.*, in a *M/FE/S* arrangement: **Fig. 1(b)** – *top*, predicated on the possibility of varying the electron density over several orders of magnitude. While Si seems to be an obvious choice for the *S* [7] the influence of the related oxides, *e.g.*, $SiO_2$ on the Si, as an IL should be carefully considered. From the continuity of electrical displacement across the *FE*-IL interface, *i.e.*, with $\varepsilon_{FE}E_{FE} = \varepsilon_{IL}E_{IL}$ and $\varepsilon_{FE} > \varepsilon_{IL}$, the $E_{FE} < E_{IL}$, implying *wasted* voltage across the IL. The electrical field in the interfacial layer ($E_{IL}$) is typically considered the "*weakest link*" in *FE* devices[1], and has also been implicated in reduced device endurance [8,9] due to the disparity in charge density - consequent to which charge compensation in the latter occurs through necessary defect generation, charge trapping, *etc*. In this context, the consideration of 2D materials is attractive, from the viewpoint of the lack of dangling bonds[10] which inhibit formation of interfacial compounds and the consequent IL.

It was then aimed to alleviate IL related issues through interfacial engineering by including single layer graphene (SLG) at the interface between the *FE* and the IL in an FTJ. The compensation of the ferroelectric polarization by the SLG minimizes the electric field in the IL underneath, consequently ensuring most of the voltage across the junction is dropped across the *FE*.

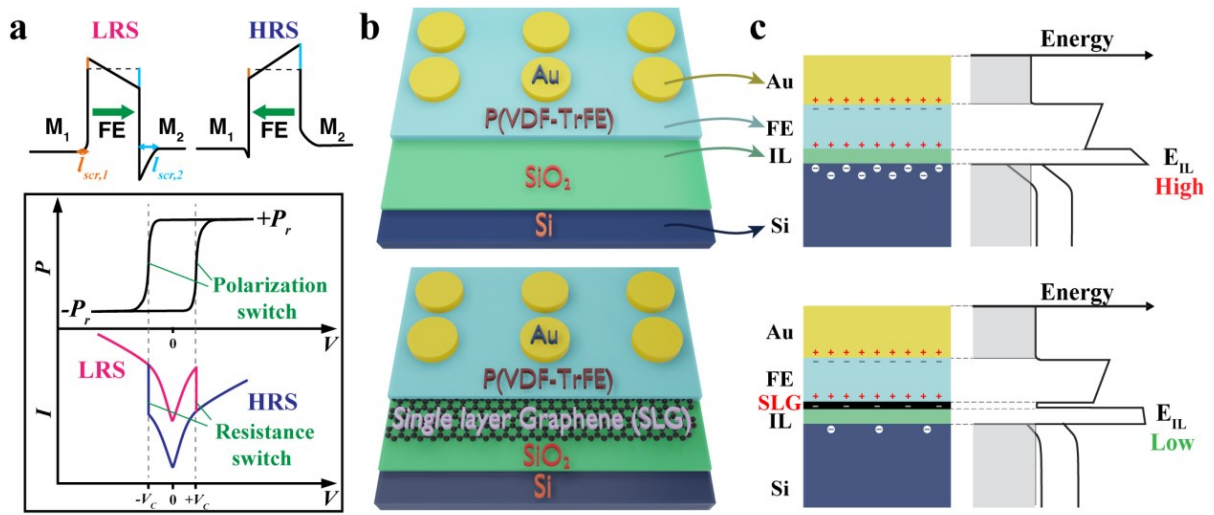

**Figure 1. Working principles and device schematic of 2-D material based ferroelectric (*FE*) tunnel junctions (FTJs). (a)** The electrical tunneling current in a FTJ (*top*) may be modulated to be in a low (/high) resistance state, *i.e.,* LRS (/HRS), depending on the direction of the *FE* polarization. The differing screening length ($l_{scr}$) values for the two electrodes, say (1) and (2) are indicated with the assumption that $l_{scr,2} > l_{scr,1}$. The consequently expected polarization (*P*) – voltage (*V*) and the current (*I*, in log scale) – *V* are indicated (*bottom figure*). **(b)** Device structure schematic of the fabricated FTJ on bare Si (*top*) and interface engineered graphene/Si surface (*bottom*). **(c)** Schematic of the cross-section and corresponding energy band structure of the FTJ system on Si (*top*), *i.e.,* the *M/FE/Si* configuration, where the charges in the FE are incompletely screened due to intervening $SiO_2$ and contribute to a large electric field across the $SiO_2$. Alternately, through the use of a 2D semiconductor (2D-*S*), such as single layer graphene, in a *M/FE/2D-S/Si* platform (*bottom*), where the 2D-*S* helps screen most of the FE charge and lowers the electric field in the $SiO_2$, a lower applied voltage would be adequate to switch the FE polarization.

For the *FE*, traditional materials such as inorganic perovskites, *e.g.,* belonging to the lead or barium titanate families, require epitaxial growth on atypical substrates for ultrathin films[11] and ultimate scaling. Further, atomic layer deposition (ALD) of $Hf_{1-x}Zr_xO$ (HZO) is non-trivial due to requiring seeding interlayers for deposition[12,13]. Consequently, P(VDF-TrFE) was selected as the *FE* layer, for this study. A particular advantage of organic ferroelectrics is the possibility of ultimate scaling to atomic dimensions, *e.g.,* through the use of a *FE* such as P(VDF-TrFE): poly-(vinylidene di-fluoride – trifluoroethylene) copolymer, where the ferroelectricity originates at the single atom level, *i.e.,* from the dipole moment between an electro-negative (/positive) F -(/H-) ion[4]. Such atomic layers may be integrated into FTJs - where high(/low) tunneling current through the *FE* in the junction is utilized to represent high(/low) states, through non-destructive resistive readout. Approaches *to date* have deployed thin *FE* P(VDF-TrFE):, placed either between metallic electrodes such as Al, Au, Pt, ITO (indium tin oxide), LSMO (Lanthanum strontium manganese oxide) – yielding reduced *TER,* or using non-silicon-based semiconductors such as Nb-STO (strontium titanate). However, as indicated earlier, if an ideal FTJ could be posited to be of the *M/FE*/2D-*S* type, we show here proof of principle towards the necessity of such a device through an intermediate *M/FE/*2D-*S/Si* structure: **Fig. 1(b) - *bottom*,** with the *net* absence of an IL - consequent to which low voltage operation is feasible. This was done, through experimental evaluation of a configuration involving the *FE* placed on graphene/Si substrate electrode. Importantly, sub - 1 V operation was obtained.

Fundamental to such implementation is the consideration of the electronic band structure of the *M/FE/Si* in comparison with the *M/FE/*2D-*S/Si*: **Fig. 1(c).** In the former case, the large $P_{FE}$ results in a corresponding large electric field across the $SiO_2$ IL, with wasted voltage. Alternately, in the latter configuration, a compensation of the $P_{FE}$ by the single layer graphene (SLG) – serving as the 2D-*S,* minimizes the electric field in the IL underneath, consequently ensuring most of the voltage is dropped across the *FE*. While the use of a Si substrate does imply superficial oxide formation, the burden of the $SiO_x$ is reduced when the SLG is used.

For this purpose, a SLG layer [14] – see section S1 of the *Supplementary Information*, was first transferred onto a Si substrate. An ultra-thin (~ 3.2 nm) P(VDF-TrFE) FE film[15] was then deposited onto *both* Si as well as *SLG/Si* substrates through the Langmuir-Blodgett (LB) technique, as schematically indicated in **Fig. 2(a)** – also see section S2 of the *Supplementary Information*.

**Results and Discussion**

**Structural analysis and verification of ferroelectric (*FE*) characteristics**: Deposited L-B P(VDF-TrFE 70:30) copolymer films/substrate were annealed at 135 °C for 30 mins in air [16] to yield the *β*-phase *FE* form [17] – inferred through x-ray diffraction on thicker samples: **Fig. 2(b)**. The subsequent probing of the surface morphology of the thin P(VDF-TrFE) film through atomic force microscopy (AFM) indicated elliptical nanorods, characteristic [18] of such a *FE* phase[19]: **Fig. 2(c)**. The presence of underlying SLG was found beneficial for smooth monolayer *FE* films, presumably due to reduced interfacial interactions [20].

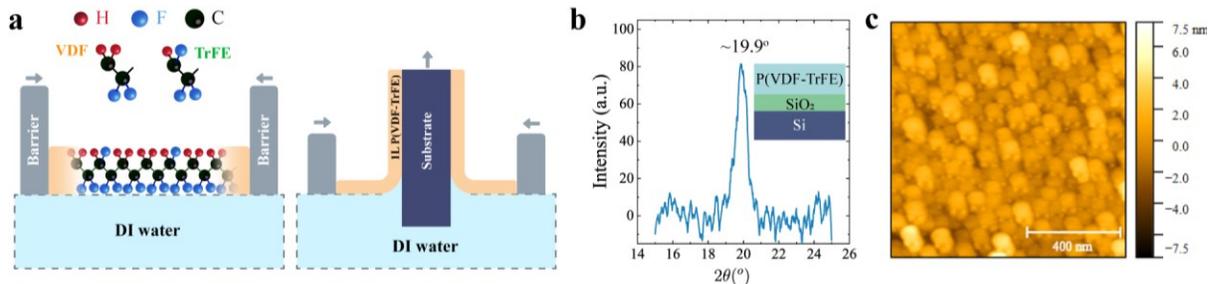

**Figure 2**. **Synthesis and structural characterization of ferroelectric (FE) P(VDF-TrFE) films deposited through the Langmuir Blodgett (L-B) technique**. (a) Sequence of steps related to the L-B deposition of the P(VDF-TrFE) films. (b) X-ray diffraction (XRD) on a thick film (~ 20 nm) of P(VDF-TrFE) showing a peak at 2θ = 19.9° corresponding to the FE *β*-phase in the polymer. (c) AFM topography image of thick film showing elliptical nanorods, characteristic of the ferroelectric *β*-phase.

PFM spectroscopy was used to indicate close to ideal 180° phase contrast characteristic [21]: **Fig. 3(a)**, as well as minima in the amplitude signal of the "butterfly" attribute: **Fig. 3(b),** at a tip voltage of +1.4 V and -1.1 V - which serve as markers of the coercive voltage [22]: $V_c$ of the order of 1.25 V (~ ½ |1.4+1.1|). Alternately, for the LB *FE* films deposited on the *SLG/Si* substrates, the *phase* and the *amplitude* plots, *i.e.,* **Figs. 3 (c)** and **3(d),** respectively, also indicate a phase reversal of ~ 180° and switching at tip voltages of +0.7 V (/-0.9 V) in the PFM amplitude scans. Hence, a $V_c$ of the order of 0.8 V (~ ½ |0.7+0.9|) was obtained. Such diminished $V_c$ – to less than 1 V, is significant and marks the considerable reduction of the influence of the underlying *IL* layer.

To illustrate the ferroelectric switching, domains were embossed on the FE film using a conductive AFM tip, the related phase and amplitude plots of which are shown in **Fig. 3(e),(f)** respectively – for SLG/Si substrates. Here, a grounded AFM tip was scanned on a 6 μm x 6 μm area with the substrate voltage ($V_{sub}$) at + 5 V, for defining a down (↓) polarization state in the *FE* film. Similarly, a smaller 2 μm x 2 μm area was switched to an up (↑) polarization state by scanning the grounded AFM tip with $V_{sub}$ at - 5 V. The *final* polarization state was read by scanning over a 6 μm x 6 μm area with $V_{sub}$ = 0 V. In **Fig. 3(e)**, the bright (/dark) phase signal is indicative of up (/down) polarization [18], induced by the AFM tip. Similarly, in **Fig. 3(f)**, the outside and inside square have comparatively high amplitude compared to the boundary where the *FE* switching occurs.

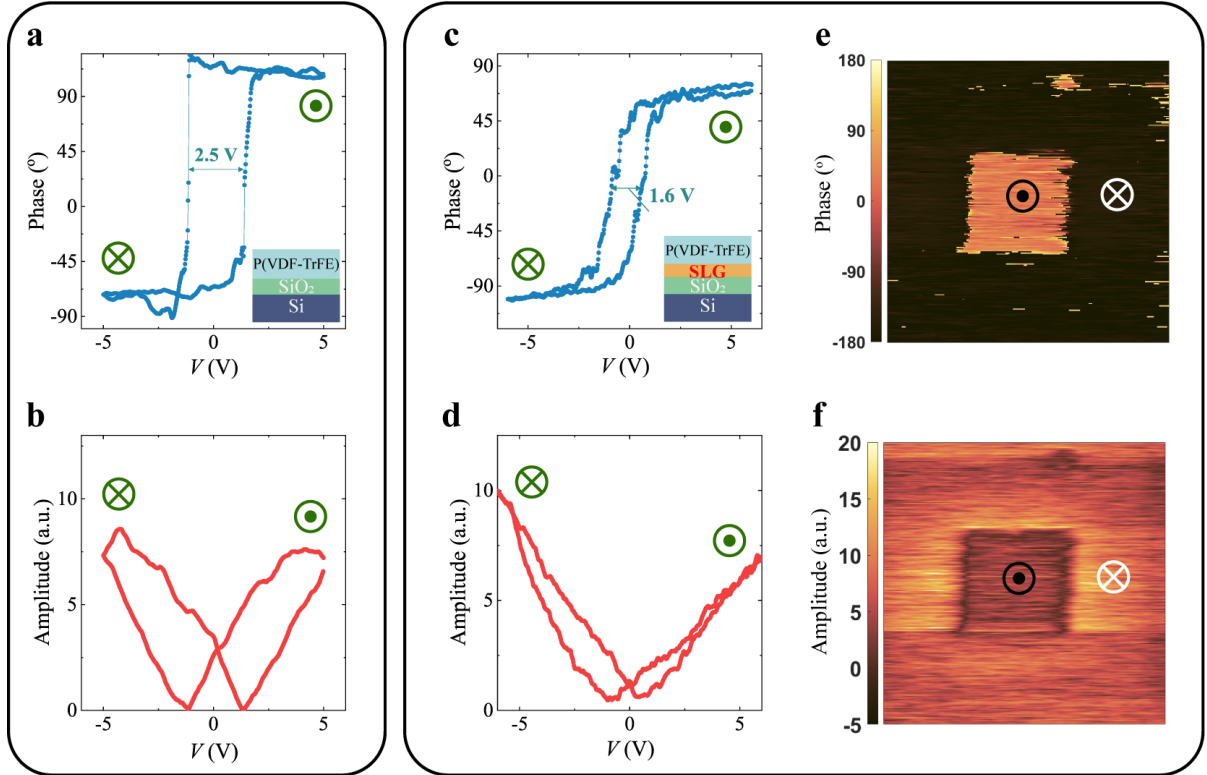

**Figure 3 | Polarization switching voltage response in P(VDF-TrFE) films deposited on silicon and SLG/silicon substrates, as inferred through piezoresponse force microscopy (PFM).** The observed **(a)** phase and **(b)** amplitude as a function of voltage on a 1L P(VDF-TrFE) film deposited on a Si substrate. The polarization switching takes place at -1.1 V and +1.4 V resulting in a $2V_C \sim 2.5$ V. The observed **(c)** phase and **(d)** amplitude as a function of voltage on a 1L P(VDF-TrFE) film deposited on a *SLG*/Si substrate. The polarization switching occurs at lower voltages, *i.e.,* - 0.9 V and + 0.7 V, resulting in a reduced $2V_C \sim 1.6$ V. The corresponding PFM **(e)** phase, and **(f)** amplitude imaging on a 1L-P(VDF-TrFE) film after switching the polarization negative in a 6 μm x 6 μm area and switching the polarization positive in the central 2 μm x 2 μm area. The • and ⊗ symbols indicate up (↑) and down (↓) polarization states, respectively. In **(e)**, the bright (/dark) phase signal is indicative of up (/down) polarization, induced by the AFM tip. In **(f)**, the outside and inside squares have higher amplitude compared to the boundary where the *FE* switching occurs.

We note that there has been discussion on the validity of PFM measurements, in terms of the influence of the mechanical strain manifested through the observed electrical signals [6], and the possibility that *FE*-like hysteresis may be obtained in non-*FE* systems. However, it was indicated [23] that the existence of an explicit $V_c$ in the DC scans *alone* could be regarded as strong evidence for intrinsic ferroelectricity and such an aspect was thoroughly verified: see section S3 of the *Supplementary Information*.

**Ferroelectric Tunnel Junction (FTJ) devices on Si and graphene/Si substrates**: Based on the attractive attribute of a lower operational voltage that may be achieved through the *FE/SLG* layer paradigm, the next step was to demonstrate the underlying principles in a device configuration. Au electrodes were patterned using shadow masks and subsequent electron-beam evaporation onto the *FE* film, for this purpose. Related current (*I*) - voltage (*V*) measurements, and the derived *TER*, with respect to the varying electrical potential on the top Au electrode, are shown in **Fig. 4** – for *both* the *M/FE/Si* and *M/FE/SLG/Si* configurations. A sudden modulation of the electrical current switching off (/on) at a voltage of ~ 1.4 V (/- 1.1 V) during the forward (/reverse) sweep was observed in the *M/FE/Si* case: **Fig. 4(a).** The modulation is over five orders of magnitude, pointing to the significant influence of the *FE* polarization. For accurate measurements of *TER*, the devices were first set(/reset) to +2 V (/-2 V) and then resistance measured in a narrow voltage range (-0.2

V to 0.2 V) – **Fig. 4(b)**. The corresponding variation of the *TER* in **Fig. 4(c)** was measured to be ~ $10^7$ %, matching previously reported highest values[21]. The high TER may be ascribed to the influence[5] of *both* the (i) barrier modulation, as well as (ii) corresponding changing of the Si from an accumulation mode to a depletion mode – affecting carrier states available for tunneling.

Similarly, in the *M/FE/ SLG /Si* case, the switching off (/on) events occurred at a voltage of ~ 0.84 V (/- 0.8 V) during the forward (/reverse) sweep: **Fig. 4(e),** with an observed decrease

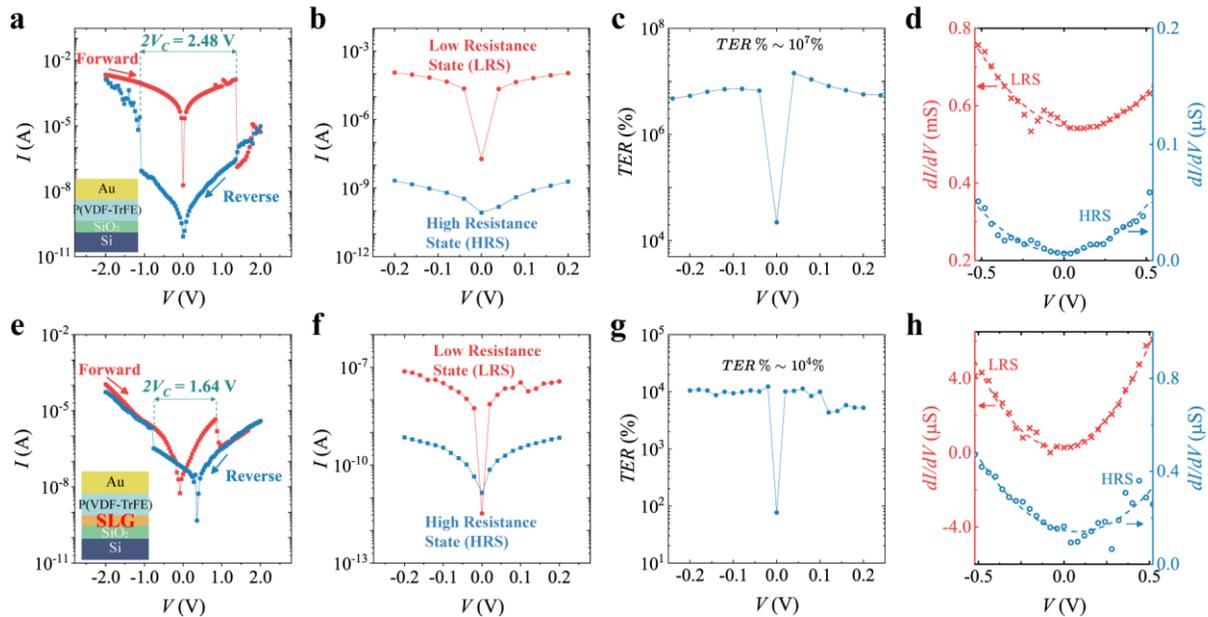

**Figure 4 | Resistive switching related to polarization modulation in P(VDF-TrFE) tunnel junctions**. *Au/FE/Si* devices – *top*. **(a)** A change from (/to) a high (/low) current state was observed in the forward (/backward) scans, at similar voltages as observed in PFM, *cf.,* Fig. 3(a) – indicating correlation between FE polarization and resistive switching phenomena, The corresponding **(b)** tunneling current (*I*) as a function of voltage (*V*) , and the **(c)** derived TER %. **(d)** The derived conductance (*dI/dV*) – voltage (*V*) characteristics.

*Au/FE//SLG/Si* devices – *bottom*. **(e)** Modulation from (/to) a high (/low) current state in the forward (/backward) scans. The **(f)** tunneling current (*I*) as a function of voltage (*V*) , **(g)** derived TER %, and **(h)** conductance characteristics.

(/increase) of the electrical current. The obtained switching voltage values, (-1.1 V and 1.4 V *i.e.* $2V_C = 2.5$ V) for *M/FE/Si* case and (-0.80 V and 0.84 V *i.e.* $2V_C = 1.64$ V) for the *M/FE/SLG/Si* case, are well matched to those obtained in the PFM scan: **Fig. 3** and clearly confirm that the resistance switching is a consequence of the *FE* character [3] of the film. The negative (/positive) shifts along the voltage axis in the forward(/backward) scans are likely due to the presence of interface/border traps, *i.e.,* that may have originated from polymer residues used in the graphene transfer process, *etc.* Similar set/reset measurements as in the *M/FE/Si* samples were carried out: **Fig. 4(f)** and the TER % was observed to be ~ $10^4$: **Fig. 4(g)** – significantly *higher* than any previously reported device with a $V_c$ < 1V.

While it is remarkable that the inclusion of *SLG* successfully reduces the operating voltages in our devices by ~ 40%, there is also indicated a substantial reduction of TER from $10^7$ to $10^4$. Given that the TER is related to the contrast of $l_{scr}$ between the two electrodes flanking the *FE* film – see **Fig. 1(a)**, the semi-metallic character of *SLG*, *i.e.,* with $l_{scr,Au}$ < $l_{scr,SLG}$ < $l_{scr,Si}$, implies reduced contrast and leads to the smaller TER.

The transport at low voltages in FTJs is based on direct tunneling with trapezoidal barriers which manifest *parabolic* conductance-voltage characteristics[24]. The related experimental measurements for the *M/FE/Si* and *M/FE/SLG/Si* configurations, *i.e.,* **Figs. 4(d)** and **4(h),** respectively – as derived from **Figs. 4(a)** and **(e)**, indicate such attributes. We analyzed the experimental *I-V* characteristics, considering tunneling barriers and related electrical potential profiles: **Fig. 5(a)** for the down (↓) and up (↑) polarization states - through a transfer matrix algorithm (TMA) based approach[25,26] – for estimating the *FE* polarization. Here, the (i) *Au/FE* barrier (of height $\phi_{Au-FE}$), (ii) *FE/SiO2 IL* interface barrier (of height $\phi_{FE-IL}$) along with the (iii) SiO$_2$ *IL*/Si interface barrier (of height $\phi_{IL,1}$ and $\phi_{IL,2}$)[27] of ~ 4.5 eV, are relevant fitting parameters. The interplay among the described potential barriers is modeled as a superposition of contributions from distinct sources. These include the potential arising from spatial variations in the conduction band minimum across the FTJ, the electrostatic potential generated by the spontaneous polarization of the *FE* layer, the polarization induced within the nonpolar dielectric layer (*IL*) due to the *FE*, and the screening charges present in the electrodes[28,29].

Here, $\varepsilon_{FE}$ (=6) [30], $t_{FE}$ (= 3.2 nm); $\varepsilon_{SiO2}$ (= 3.9), $t_{SiO2}$ (=1.3 nm) indicate the respective dielectric permittivity and thickness of the *FE* and *IL*. Under a voltage bias (say, $V_b$), the TMA yields an energy (*E*) dependent transmission probability: $T(E, V_b)$, which was used for estimating [31] the related tunneling current: $I(V_b)$, through the relation:

$$I(V_b) = A \frac{e}{2\pi\hbar} \int_0^\infty T(E, V_b)[f(E) - f(E')]dE \quad (1)$$

Here, *A* is the top electrode area, *e* is the elementary unit of electron charge, $\hbar$ the reduced Planck's constant, $E' = E + eV_b$, and $f(E)$ is the relevant supply function of the top and bottom electrode/s - a measure of the flux of electrons through the barrier, proportional to the *k-vector* dependent carrier velocity ($v$), the energy dependent density of states: $DOS$ ($E$), and the state occupation probability (through the Fermi-Dirac distribution function: $f_{F-D}$), defined by:

$$f(k) = \int_0^\infty v \cdot DOS(k) f_{F-D}(E) dk = \frac{1}{\hbar} \int_0^\infty DOS(E) f_{F-D}(E) dE \equiv f(E) \qquad (2)$$

The *FE* polarization induces in the Si either accumulation *or* depletion of carriers, varying the $(DOS)_{Si}$ and hence modulating the supply function by $\sim 2 \cdot 10^4$, as further discussed in Section 4 of the *Supplementary Information*. Through experimental knowledge of the left hand side of Eqn. (2), the fits to $\phi_{Au-FE}$ and $\phi_{FE-IL}$ along with the $\phi_{IL,1}$ and $\phi_{IL,2}$, were determined, and have been indicated in conduction band (CB) related potential profiles, for both polarization directions in the *FE* P(VDF-TrFE) on Si in **Fig. 5(a)** along with the related modeled *I-V* characteristics in **Fig. 5(b)**.

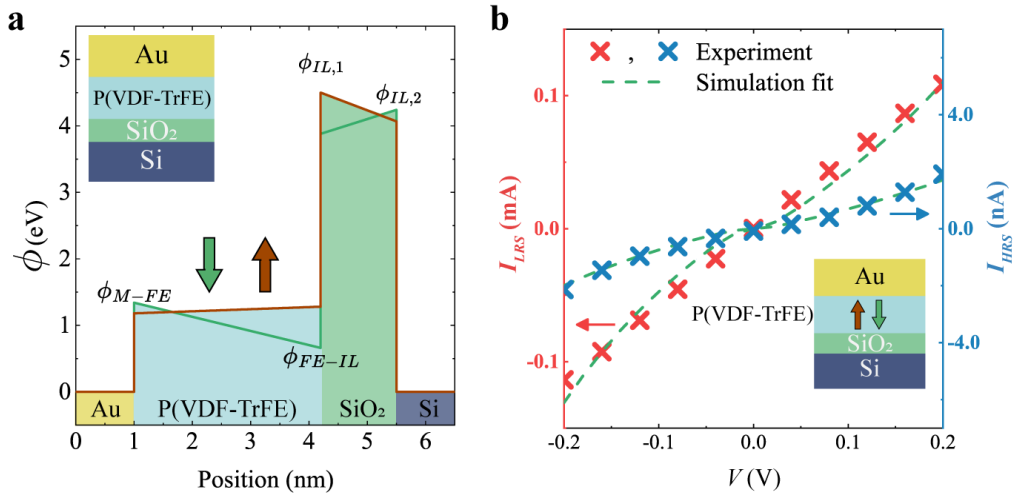

**Figure 5 | Modeling tunneling current through trapezoidal tunneling barriers, for up (↑) and down (↓) polarization states, through a transfer matrix algorithm (TMA) approach.** **(a)** Using $\phi_{Au-FE,}$ for ↑ (/↓) configuration as 1.2 eV (/1.3 eV) and $\phi_{FE-IL}$ for ↑ (/↓) configuration as 1.3 eV (/0.7 eV) related to the high (/low)-resistance states: *HRS* (/*LRS*), respectively, yielded **(b)** an excellent fit to the observed $I - V_b$ characteristics.

The related barrier height modulation, *i.e.*, $|\phi_{M-FE,} \uparrow - \phi_{M-FE,} \downarrow|$ related to the *upper* Au electrode, arising from polarization switching, is $\sim 0.1$ eV, and on the *bottom* Si electrode is $\sim 0.6$ eV. The effect of the larger $l_{scr,Si}$ is implicit in the higher modulation of the *FE* barrier on the Si side. Assuming a screening length of $\sim 0.07$ nm for Au[32] and $\sim 0.41$ nm for Si[33] and combining them with the barrier modulations [34], we estimate a polarization of 6.0 - 8.0 µC/cm$^2$ for the P(VDF-TrFE) films, which seems a typical value[35].

The obtained results, from our work are superior with respect to two crucial key performance indicators, *i.e.,* the TER and the coercive voltage ($2V_c$), in comparison to reports from literature [36] – as indicated in **Fig. 6.** We have obtained a TER (%) of $10^7$ and $\sim 2 \cdot 10^4$ along with a $V_c$ of ~ 1.3 V and ~ 0.8 V for the *Au/FE/Si* and *Au/FE/SLG/Si* devices, respectively. As indicated in the introductory sections, an operating voltage of less than 1 V would be crucial in breaking down a barrier for encouraging adoption of *FE* technology and this has been achieved in our work.

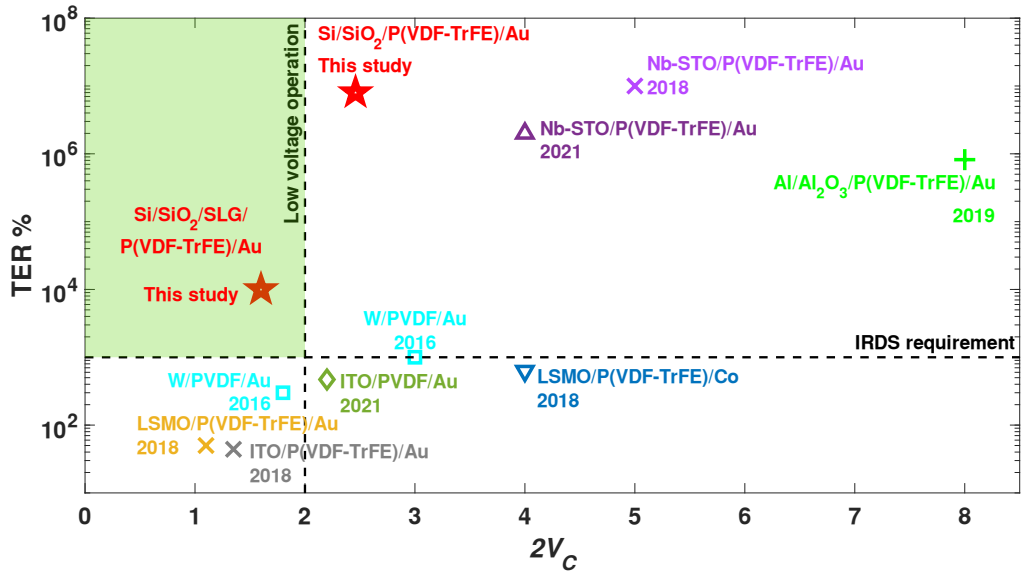

**Figure 6 | Benchmark of device performance** Benchmark of PVDF based ferroelectric tunnel junctions comparing *TER*% as a function of coercive voltages in the literature. The green highlighted region represents the ideal range of device performance, where the lowest acceptable TER% is dictated by IRDS requirements and the highest acceptable coercive voltage is dictated by compatibility requirements with logic in logic-in-memory applications.
[4,21,41,42,43]

Further, a TER% of at least $10^3$ was recommended in the *International Roadmap for Devices and Systems* (IRDS)[37], and has been demonstrated here.

Further, our results constitute the first reported instance of P(VDF-TrFE) based FTJs fabricated on Si based systems through the synergetic incorporation of 2D materials. Our work, based on low temperature synthesis and processing coupled with the lower operating voltage/s, provides further impetus for the integration of *FE* materials and systems into back-end-of-line (BEOL) implementations. Given recent imperative for the incorporation of 2D materials [38] in *FE* devices[39], our work would further motivate the development of large scale assembly of L-B films[15].

**Methods.**

**Materials Synthesis:**

Ferroelectric layer deposition. The P(VDF-TrFE) copolymer (from PolyK Technologies), constituted of 70% vinylidene fluoride (VDF) and 30% trifluoroethyene, was dissolved in cyclopentanone with a concentration of 0.1 wt%. The solution was dispersed on the surface of deionized water in a KSV Nima Langmuir trough with an associated dipper mechanism. The surface pressure was continuously monitored, and the surface was compressed to ~ 5.3 mN/m, as measured using a Wilhelmy plate[40]. Subsequently, the Silicon and single layer graphene (SLG) coated Silicon substrates – *see next section,* were raised through the surface polymer film, depositing a single layer of the P(VDF-TrFE), following which the substrate was allowed to dry. The associated deposition technique – termed the Langmuir Blodgett (L-B) methodology, could be repeated multiple times to yield thicker films.

Single layer graphene (SLG) was synthesized using chemical vapor deposition (CVD) in a quartz tube reactor on 25 μm thick copper foils (MTI Corp) over 2 h, using a mixture of methane (2 sccm) and hydrogen (15 sccm) gas as the precursors for the deposition. To transfer the graphene onto Si substrates, the graphene covered Cu foil was coated with a 3 wt% P(VDF-TrFE) film using spin-coating. Next, Oxygen plasma reactive ion etching (Trion RIE), was used to remove the excess graphene on the other side of the foil. The foil was then floated in an ammonium persulfate (0.1 M) solution to etch off the Cu and subsequently rinsed with DI water. The graphene/polymer composite was then transferred onto the substrate. The P(VDF-TrFE) was dissolved using acetone for 4 h followed by IPA rinse and $N_2$ blow dry after which the Langmuir Blodgett deposition was carried out to deposit the ultrathin ferroelectric film.

**Characterization by Atomic Force Microscopy (AFM) and Piezoresponse Force Microscopy (PFM)** was performed at room temperature in a Park NX 20 AFM system. Topography AFM images were taken using the non-contact mode using commercial silicon tips (MikroMasch). PFM measurements were performed using silicon tips coated with Cr/Pt (obtained from Budget Sensors). Here, an AC voltage of ~ 1 V (at ~12 kHz frequency) superimposed on a slower varying (of period ~ 1 s) triangular sweep voltage (DC) in a range of ± 1 V to ± 5 V, was applied to the tip scanning the device/s and the resulting measured current, with respect to the phase and amplitude, was monitored as a function of voltage bias.

**Film thickness** was measured using (i)AFM, (ii) nanoindentation, and (iii) ellipsometry. Film thickness was measured across a fabricated step edge by AFM. On the average, a thickness of one P(VDF-TrFE) monolayer was determined to be ~ 3.2 nm. Alternately, iMicro (from KLA Inc.) equipped with a Berkovich tip (Type TB30524) was used for nanoindentation based measurement of the film thickness, comparing a *bare* silicon sample and one coated with 1L-P(VDF-TrFE) to give a thickness ~ 2.9 nm. Further, ellipsometry (J.A. Woollam M-2000D) was also used to fit the film thickness, in the Cauchy mode, to verify that one monolayer of the FE film yielded a thickness of ~ 3.3 nm and the thickness of the $SiO_2$ layer to be 1.3 nm. An average thickness of 3.2 nm was assumed for the 1L-P(VDF-TrFE) film.

**Electrical measurements** were performed on the samples (with 50 nm thick gold electrode contacts of 150 μm diameter deposited using e-beam evaporation and patterned using a shadow mask) in a shielded probe station using an Agilent B1500A semiconductor device parameter analyzer. *Soft* evaporation parameters (vacuum ~ $5 \cdot 10^{-7}$ Torr, deposition time < 10 mins, distance between gold source and sample ~ 50 cm) were used to minimize damage to the film.

For current (*I*) – voltage (*V*) measurements, the substrate was grounded and bias voltage was applied to the top Au electrodes. The voltage was swept from positive to negative bias and back at a sweep rate of ~120 mV/s in steps of 40 mV. For TER measurements, the device was biased at a large positive(/negative) voltage = +2 V(-2 V) to switch the ferroelectric film into a given polarization state. Then, a small sweep of voltage from -0.2 to 0.2 V in steps of 20 mV was applied and current measured. It was ensured that trapping charges did not shift the *I-V* curve minima away from 0 V, through such measurement protocols.

**Data Availability.**

The data that support the plots within this paper and other findings of this study are available from the corresponding author upon reasonable request.

15. Ducharme, S., Reece, T. J., Othon, C. M. & Rannow, R. K. Ferroelectric polymer Langmuir-Blodgett films for nonvolatile memory applications. *IEEE Transcations Devices Mater. Reliab.* **5**, 720 (2005).
16. Zhao, C. *et al.* Enhanced ferroelectric properties of P(VDF-TrFE) thin film on singlelayer graphene simply adjusted by crystallization condition. *Mater. Today Energy* **20**, 100678 (2021).
17. Yin, Z., Tian, B., Zhu, Q. & Duan, C. Characterization and Application of PVDF and its Copolymer Films Prepared by Spin-Coating and Langmuir–Blodgett Method. *Polymers (Basel).* **11**, 2033 (2019).
18. Qian, J. *et al.* Temperature dependence of piezo- and ferroelectricity in ultrathin P(VDF–TrFE) films. *RSC Adv.* **8**, 29164 (2018).
19. Feng, T. *et al.* Temperature Control of P(VDF-TrFE) Copolymer Thin Films. *Integr. Ferroelectr.* **141**, 187–194 (2013).
20. Xia, W. *et al.* Epitaxy of Ferroelectric P(VDF-TrFE) Films via Removable PTFE Templates and Its Application in Semiconducting/Ferroelectric Blend Resistive Memory. *ACS Appl. Mater. Interfaces* **9**, 12130 (2017).
21. Majumdar, S., Chen, B., Qin, Q. H., Majumdar, H. S. & van Dijken, S. Electrode dependence of tunneling electroresistance and switching stability in organic ferroelectric P(VDF-TrFE)- based Tunnel junctions. *Adv. Funct. Mater.* **28**, 1703273 (2018).
22. Guan, Z. *et al.* Identifying intrinsic ferroelectricity of thin film with piezoresponse force microscopy. *AIP Adv.* **7**, 095116 (2017).
23. Guan, Z. *et al.* Identifying intrinsic ferroelectricity of thin film with piezoresponse force microscopy. *AIP Adv.* **7**, 1–8 (2017).
24. Brinkman, W. F., Dynes, R. C. & Rowell, J. M. Tunneling Conductance of Asymmetrical Barriers. *J. Appled Phys.* **41**, 1915 (1970).
25. Eshraghi, K. & Bandaru, P. R. Enhancement of photoelectron emission efficiency from quantum dot solids, through electrical field biasing of interfaces. *Appl. Phys. Lett.* **118**, 263104 (2021).
26. Eshraghi, K., Natani, S. & Bandaru, P. R. Modeling electronic conduction in quantum dot constituted assemblies coupled to metallic electrodes. *Appl. Phys. Lett.* **122**, (2023).
27. DiStefano, T. H. & Eastman, D. E. Photoemission measurements of the valence levels of
14

**Acknowledgements**

**Author Contributions**
S. Natani (S.N.) and P.R. Bandaru (P.R.B.) conceived the ideas and experiments, which were mainly performed by S.N.. Li Cheng (L.C.) contributed in the growth of single layer graphene. Pranjali Khajanji (P.K.) and Wade Shipley (W.S.) contributed to the L-B film deposition. Kassra Eshraghi (K.E.) and S.N. contributed to the simulation study. All the authors contributed to the analysis and interpretation of the results, and the writing of the manuscript.

**Competing Interests** The authors declare no competing interests.

**Corresponding author** Correspondence to P.R. Bandaru

# Supplementary information

**Content:**

**S1. Synthesis and characterization of single layer graphene (SLG)**

**S2. Evaluation of the P(VDF-TrFE) film morphology and thickness**

**S3. Interpretation of the piezo-force microscopy (PFM) measurements in terms of ferroelectric (*FE*) characteristics**

**S4. The influence of ferroelectric (*FE*) polarization on the electrode supply function/s**

**S5. Estimation of the *FE* polarization**

**S6. Alternate device for 1L-P(VDF-TrFE) on SLG/Si**

**S1. Synthesis and characterization of single layer graphene (SLG):** The graphene was synthesized through chemical vapor deposition (CVD) on copper foil [1] placed in quartz tube, using methane precursor gas at a temperature of ~ 1050 ºC. Prior annealing of the foils for 30 minutes (with 15 sccm of hydrogen and 5 sccm of argon gases) was found to be useful for higher quality graphene. The flow rate of the methane was used for tuning the number of graphene layers, *i.e.,* SLG was obtained at a methane flow rate of 1 sccm for 3 hours, while few-layer graphene (FLG) was obtained with larger flow rates[2], say ~ 4 sccm. The SLG was subsequently transferred onto a silicon substrate.

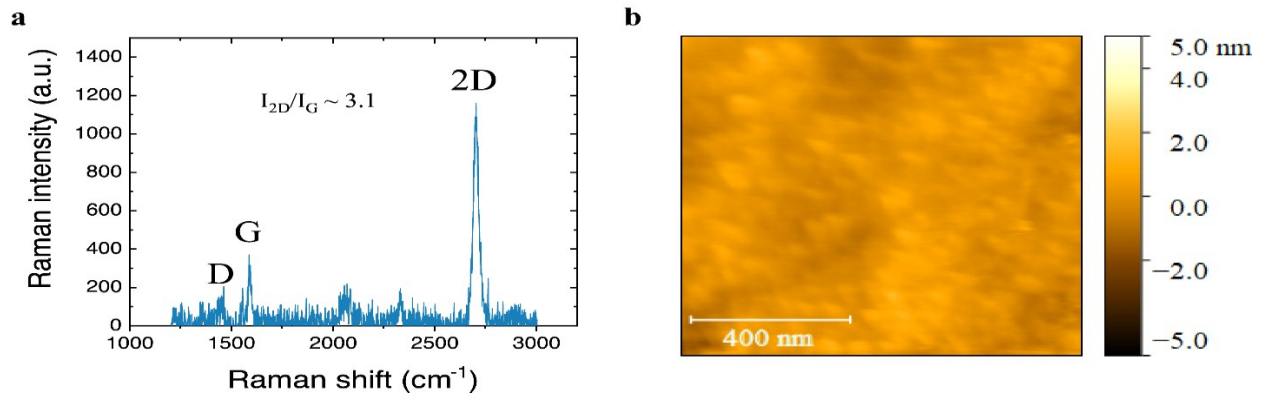

**Figure S1 :** (a) Raman spectra of *transferred* graphene on Silicon substrates indicating SLG characteristic, with a 2D-/G-peak intensity ratio greater than two. (b) AFM topography scan of the SLG transferred onto Si indicates a smooth surface with no polymer residues.

Raman spectroscopy (Renishaw inVia™ confocal Raman microscope) coupled with Atomic Force Microscopy (AFM, Park NX20) were used to monitor the quality of the graphene. Fig. S1(a) shows a typical Raman spectra of the transferred graphene, indicating the three characteristics peaks[3], *i.e.*, the 2D peak (~ 2780 cm$^{-1}$), G peak (~ 1540 cm$^{-1}$) and the D peak (~1350 cm$^{-1}$). Broadly, it has been indicated [3] that a peak intensity (*I*) ratio, *i.e.*, $I_{2D}/I_G$ greater than 2 implies SLG, as revealed through the spectra. AFM measurements on the surface: Fig S1(b), indicates a smooth film (RMS roughness ~ 0.4 nm) ensuring no polymer residues or related surface defects.

**S2. Evaluation of the P(VDF-TrFE) morphology and film thickness**. The deployed FE: P(VDF-TrFe 70:30) copolymer, was constituted from 70% VDF and 30% TrFE, the latter typically used for the crystallization from the melt into a polar *FE*, of the *β*-phase variety[4]. Such a specific phase, considered characteristic of an FE film, was obtained through annealing of the film at 135 °C for 30 mins in air [5]. Three different methods, *i.e.*, (i) Atomic Force Microscopy (AFM), (ii) ellipsometry, and (iii) nano-indentation, were utilized to confirm the *one-layer* characteristic of the L-B deposited P(VDF-TrFE) polymer film. An AFM (Park NX 20) scan of the step edge between the film and the substrate indicated a thickness of ~ 3.2 nm: Figure S2 (a). Alternately, ellipsometry (J.A. Woollam M-2000D Ellipsometer) related to the polymer film deposited on the SiO$_2$/Si substrate, while adapting a Cauchy model, was used to estimate the thickness of the P(VDF-TrFE) as ~ 3.3 nm along with that of the SiO$_2$ as ~ 1.3 nm: Figure S2 (b).

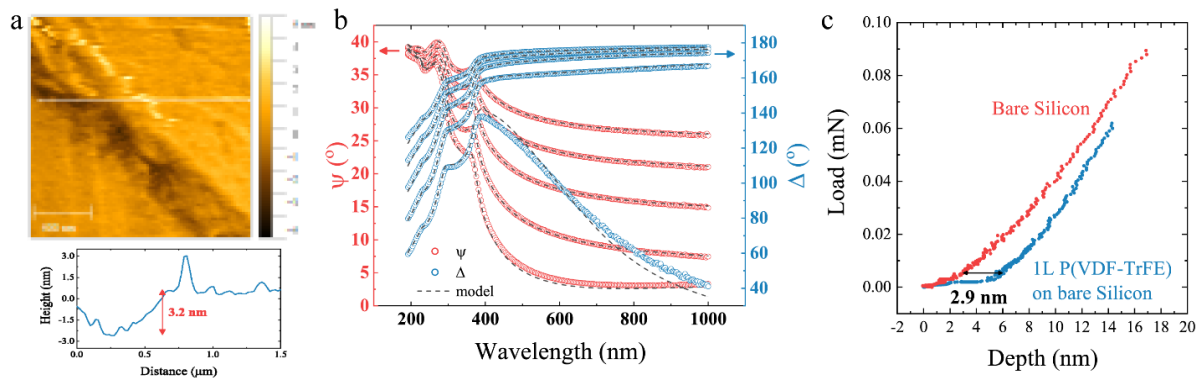

Figure S2 : **1L P(VDF-TrFE) film thickness measurement. (a)** *Top* : AFM topography scan of a fabricated step edge of 1L P(VDF-TrFE) on Si. *Bottom*: Line scan on the AFM image indicating a step edge height ~ 3.2 nm. **(b)** Ellipsometry data and model fit for 1L P(VDF-TrFE) on Si, at incidence angles 55º, 60º, 65º, 70º, and 75º. The fitting suggests a film thickness of ~ 3.32 ± 0.05 nm. **(c)** Nano-indentation characteristics comparing a bare Si substrate with 1L P(VDF-TrFE) on the Si indicates a polymer film thickness of ~ 2.9 nm.

Additionally, the variation of the indentation depth as a function of the applied load was monitored for *both* a control sample of Si as well as a sample of 1L P(VDF-TrFE) polymer deposited on silicon. The nanoindenter tip (Berkovich tip in the iMicro instrument, KLA Inc.) would be expected to easily pierce through the soft polymer layer to the Si substrate. Hence, there would be a delayed response compared to the bare silicon, whereby the horizontal shift in the load-depth characteristics would be equivalent to the polymer film thickness. As seen in Fig. S2(c), the polymer thickness was then estimated to be ~ 2.9 nm.

**S3. Interpretation of the piezo-force microscopy (PFM) measurements in terms of ferroelectric *(FE)* characteristics.** There has been much discussion on the validity of PFM measurements, in terms of the manifestation of the mechanical strain through the observed electrical signals [6] bringing forth the possibility that *FE-like* hysteresis may even be obtained in non-*FE* systems. However, it was indicated [7] that the existence of an explicit coercive voltage ($V_c$) in the DC scans *alone* could be regarded as strong evidence for intrinsic ferroelectricity. Such a DC scan measurement is shown in Fig. S3. It is noted, as indicated in the *Methods* section, that the measurement involves the application, to the tip scanning the surface of an AC voltage of ~ 1 V (at ~12 kHz frequency) superimposed on a slower varying (of period ~ 1 s) triangular sweep voltage (DC) in a range of ± 1 V to ± 5 V. Further, the resulting measured current, with respect to the phase and amplitude, was monitored as a function of voltage bias.

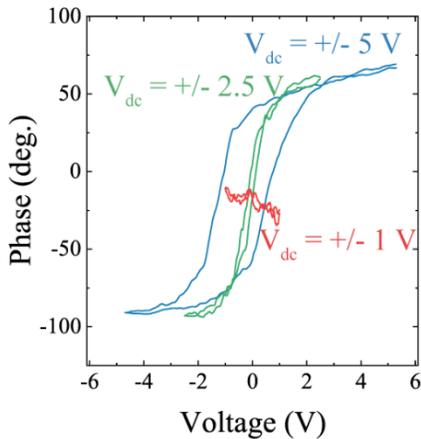

Figure S3 : Piezoelectric force microscopy (PFM) - voltage scans on the P(VDF-TrFE) polymer taken at different DC bias sweep conditions (on the PFM probe tip). There is a clear lack of switching characteristics and related collapse of the PFM loop, as the DC bias range is reduced to below the coercive voltage, *i.e.,* at ± 1 V, in comparison to the sweep ranges at higher values.

Here, it was seen that as the applied DC bias is reduced that the width of the phase-voltage characteristics decreases. As the voltage is finally reduced to below the $V_c$, there is no evidence for switching – implying true FE characteristics, as well as true $V_c$. Such an aspect would not be observed if charge injection, electrostatic effects, or mechanical strains were involved.

**S4. The influence of ferroelectric (*FE*) polarization on the electrode supply function/s.** While a polarization change in the ferroelectric (*FE*) has marginal effect on the electron supply in the metal, a large shift in the Fermi energy ($E_F$) would be obtained in the Si. Consequently, the electron supply function: *f(k)* – see Eqn. 3, in the *main text*, of the top and bottom electrode/s would be significantly modified by the *FE* polarization. We have previously defined, in the main text, the *f(k)* as a measure of the flux of electrons, from the supplying electrode, through the barrier - being proportional to the velocity as well as the density of states (*DOS*) and the Fermi-Dirac distribution function: $f_{F-D}$. There would then be expected to be a significant modulation of the *DOS* of the Si, through the *FE* polarization. For instance, in a *p-type* Si substrate, with up (↑) and down (↓) polarization in the *FE*: <u>Fig. S4(a),</u> the $E_F$ is shifted into (/away from) the bulk valence band (VB) of the Si, corresponding to more(/less) induced hole ($h^+$) carriers.

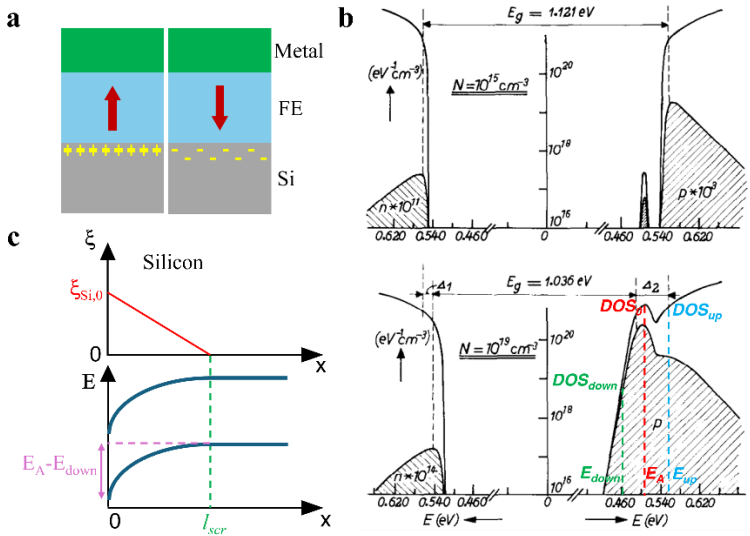

Figure S4 : **Supply function modelling for degenerately doped Si. (a)** Schematic depicting up (↑) and down (↓) polarization in the *FE* and subsequent change in charge density for *p-type* Si. **(b)** adapted from [8]. For Si, a larger (/smaller) band gap of ~ 1.121 eV (/~ 1.036 eV) was noted for non-degenerate (/degenerately) doped samples – as in the figure/s at the *top* (/*bottom*). The smaller band gap arises from the merging of the acceptor levels/band with the valence band of bulk Si. **(c)** Schematic distribution of electric field (E) in Si due to the *FE* (*top*) and related band bending at the interface due to the polarization (*bottom*).

We note that in a *p-type* semiconductor, the $E_F$ resides in the acceptor level and the *DOS(E)* in the acceptor band is described through[8]:

$$DOS(E) = DOS_0 e^{-\frac{(E_A-E)^2}{2\sigma_e^2}} \tag{S1}$$

Here, $E_A$ is the acceptor energy level, $\sigma_e$ - the effective standard deviation related to the acceptor energy level distribution and $DOS_0$ is the density of states when $E=E_A$.

In this study, due to using a degenerately doped *p-type* Si sample (doped at $\sim 10^{20}$ cm$^{-3}$), the acceptor level band combines with the Si VB, during the up (↑) *FE* polarization state, yielding a reduced energy gap for the Si – as seen in Fig. S4(b) – *bottom* [8]. Consequently, we assume that in Si, the $DOS(E) \propto (E_V - E)^{\frac{1}{2}}$, where $E_V$ is the VB energy. This implies that for small shifts in the $E_F$ - due to the *up* (↑) *FE* polarization state, the *DOS* will not experience large modulation, *i.e.*, $DOS(E_{up}) \sim DOS(E_A)$. However, with the *down* (↓) *FE* polarization state, *i.e.*, $DOS(E_{down})$, we need to use Eqn. S1. Then, the ratio of the respective *DOS* would be:

$$\frac{DOS(E_{up})}{DOS(E_{down})} \sim \frac{DOS_0}{DOS_0 \, e^{-(E_A - E_{down})^2 / 2\sigma_e^2}} \sim e^{(E_A - E_{down})^2 / 2\sigma_e^2} \qquad (S2)$$

Further, if it is assumed that the electric field: $\xi_{Si}(x)$ at the Si interface, *i.e.*, $\xi_{Si,0}$, due to the *FE* polarization is screened in the Si over a length ($l_{scr,Si}$): Fig. S4(c), the shift in the $E_A$ in the *down* (↓) *FE* polarization state, is estimated through:

$$E_A - E_{down} = e \int_0^\infty \xi_{Si}(x)dx = \int_0^{l_{scr,Si}} \xi_{Si}(x)dx = \frac{e\xi_{Si,0} l_{scr,Si}}{2} \qquad (S3)$$

At the *FE*/Si interface, with $P_r + \epsilon_0 \epsilon_{FE} \xi_{depol} = \epsilon_0 \epsilon_{Si} \xi_{Si,0}$. The depolarization field: $\epsilon_0 \epsilon_{FE} \xi_{depol}$ was estimated from [9]:

$$\epsilon_0 \epsilon_{FE} \xi_{depol} = -(P - Q_s) \qquad (S4a)$$

$$Q_s = \frac{P t_{FE}}{\epsilon_{FE} \left( \frac{l_{scr,Au}}{\epsilon_{Au}} + \frac{l_{scr,Si}}{\epsilon_{Si}} \right) + t_{FE}} \qquad (S4b)$$

The $Q_s$ is the related screening charge. Further, $t_{FE} = 3.2\ nm$, $\epsilon_{FE} = 6$, $\epsilon_{Au} = 8.64$, $l_{scr,Au} = 0.07\ nm$, $\epsilon_{Si} = 11.7$, and $l_{scr,Si} = 0.41\ nm$ [8]. The magnitude of the $\epsilon_0\epsilon_{FE}\xi_{depol}$ was estimated as ~ 0.08 $P$ and neglected. Hence, Eqn. S2 is modified to:

$$\frac{DOS_{up}}{DOS_{down}} \sim e^{-\frac{\left(\frac{el_{scr,Si}P_r}{2\epsilon_{Si}}\right)^2}{2\sigma_e^2}} \tag{S5}$$

Assuming a $P_r = 6.0\ \mu C/cm^2$ [10] – see next in Section S5, and a $\sigma_e = 0.0264\ eV$ [8] the $\frac{DOS_{up}}{DOS_{down}}$ is ~ 2·10$^4$ – as indicated in the main text. Given the direct dependence of the supply function: $f(k)$ to the *DOS* – see Eqn. (3) in the *main text*, a similar ratio was assumed for the relative supply function/s of the semiconductor electrode/s for the *FE* in the ↑ and ↓ configurations.

**S5. Estimation of the *FE* polarization.** The polarization: *P,* of the P(VDF-TrFE) film was estimated from the barrier heights obtained through our fitting of the experimental *I-V* curves. The change in barrier height at the Au/*FE* or Si/*FE* interface due to polarization switching, *i.e.,* see **Fig. 5** in the *main text,* where the $\phi_{Au\text{-}FE,}$ for ↑ (/↓) configuration was determined to be 1.2 eV (/1.3 eV) and $\phi_{FE\text{-}IL}$ for ↑ (/↓) configuration was determined to be 1.3 eV (/0.7 eV), following the work by Pantel*, et. al*[9]. Here, the change in barrier height ($2\Delta\varphi_i$) at the $i^{th}$ interface ($i = Au/FE$ or $FE/IL$):

$$\Delta\varphi_i = \frac{l_{scr,i}Q_s}{\epsilon_0\epsilon_{M,i}}e \tag{S6}$$

The $l_{scr,i}$ and $\varepsilon_{M,i}$ are the respective screening length and dielectric constant of the at the interfaces,. Now, from $2\varphi_{Au/FE} = 0.1\ eV$ and $2\varphi_{FE/IL} = 0.6\ eV$ and using Eqn. S6 and Eqn. S4(b), we estimate a $P \sim 6 - 8\ \mu F/cm^2$, the two limits corresponding to the two values of the barrier height.

## S6. Alternate device for 1L-P(VDF-TrFE) on SLG/Si

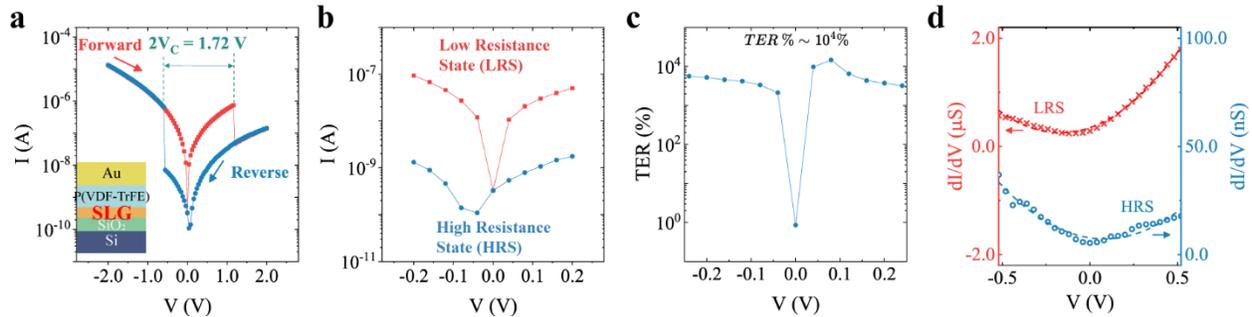

Figure S5 : **Alternate device for FTJ based on 1L-P(VDF-TrFE) on SLG/Si. (a)** I-V sweep of FTJ showing ~ 2 order of magnitude resistance switching at voltages -0.5 and +1.2 adding up to a $2V_C$ = 1.7 V. **(b),(c)** The corresponding tunneling current as a function of voltage over a narrow range and derived measurement of the TER for the device. **(d)** The conductance ($dI/dV$) – voltage ($V$) characteristics.